\documentclass[preprint,prl,amsmath,amssymb]{revtex4-1}

\usepackage{graphicx}
\usepackage{mathptmx}

\usepackage{color}
\usepackage{ulem}

\begin{document}

\title{Restoring a free-standing character of graphene on Ru(0001)}

\author{Elena Voloshina,$^{1,}$\footnote{Corresponding author. E-mail: Elena.Voloshina@hu-berlin.de} Nikolai Berdunov,$^2$ and Yuriy Dedkov,$^{2,}$\footnote{Corresponding author. E-mail: Yuriy.Dedkov@specs.com, Yuriy.Dedkov@icloud.com}}

\affiliation{\mbox{$^1$Humboldt-Universit\"at zu Berlin, Institut f\"ur Chemie, 10099 Berlin, Germany}}

\affiliation{\mbox{$^2$SPECS Surface Nano Analysis GmbH, Voltastra\ss e 5, 13355 Berlin, Germany}}

\date{\today}

\begin{abstract} 
Realization of a free-standing graphene is always a demanding task. Here we use scanning probe microscopy and spectroscopy to study the crystallographic structure and electronic properties of the uniform free-standing graphene layers obtained by intercalation of oxygen monolayer in the ``strongly'' bonded graphene/Ru(0001) interface. Spectroscopic data show that such graphene layer is heavily $p$-doped with the Dirac point located at $552$\,meV above the Fermi level. Experimental data are understood within DFT and the observed effects are in good agreement with the theoretical data.
\end{abstract}

\maketitle

\section*{Introduction}

The physics and chemistry of graphene (gr) on metals is one of the exciting fields of surface science~\cite{Tontegode:1991ts,Wintterlin:2009,Dedkov:2012book,Batzill:2012,Dedkov:2015kp}. These extensive studies allowed to understand the main interaction mechanisms at the graphene-metal interfaces and connect them with the changes in the electronic structure of graphene~\cite{Wintterlin:2009,Voloshina:2012c,Voloshina:2014jl}. Even in the earlier studies of the graphene/metal interfaces the aim was not only to obtain information about interface itself (its geometry, bonding mechanism, its electronic structure), but to try modifying the properties of interface and of graphene. The latter can be realized, \textit{e.\,g.}, via adsorption of different species on top of the system or via intercalation of different materials between a graphene layer and a metallic substrate~\cite{Tontegode:1991ts}. For the intercalation-like systems the most exciting cases include the intercalation of big molecules, like C$_{60}$~\cite{Rutkov:1995,Shikin:2000a}, and molecules of gases, like CO~\cite{Granas:2013tl,Jin:2014cl} or oxygen~\cite{Zhang:2009qqq,Sutter:2010a,Larciprete:2012aaa,Granas:2012cf}. Initial experiments on oxygen intercalation were performed for the nearly free-standing graphene on Ir(111) (islands or complete layer)~\cite{Granas:2012cf,Larciprete:2012aaa} or for \textit{strongly-interacting} graphene/Ru(0001) where graphene has an incomplete monolayer coverage that allows using a low partial pressure of oxygen gas~\cite{Zhang:2009qqq,Sutter:2010a,Jin:2012ki,Sutter:2013kw,Jang:2013cn}. Recently, it was demonstrated that oxygen can be also intercalated under full layer of graphene on Ru(0001) at the partial pressure of oxygen less than 1\,mbar and 150$^\circ$\,C~\cite{Dong:2015ig}. Oxygen intercalation under graphene on Ir(111) or Ru(0001) leads to the electronic decoupling of graphene from substrate with the corresponding $p$-doping of graphene as was found by X-ray photoelectron spectroscopy (XPS)~\cite{Larciprete:2012aaa,Granas:2012cf,Dong:2015ig}. Angle-resolved photoelectron spectroscopy (ARPES) data give for graphene/O/Ir(111) the position of the Dirac point ($E_D$) at $0.64$\,eV above the Fermi level ($E_F$)~\cite{Larciprete:2012aaa}. For graphene/O/Ru(0001) situation is controversial: the first ARPES experiments with a rather poor energy resolution give a $p$-doping of graphene in this system with the position of $E_D$ at $\approx0.5$\,eV above $E_F$~\cite{Sutter:2010a}, whereas the local STM/STS results give this position at $0.48$\,eV below $E_F$~\cite{Jang:2013cn}, that can be considered as an artefact, because the interpretation of STS data for the graphene/metal systems is not a straightforward task.

In the present work we demonstrate the possibility to form the graphene/O/Ru(0001) intercalation system at the relatively low partial oxygen gas pressure and the elevated temperature for the complete initial graphene coverage on Ru(0001). The formed system of very high quality demonstrate the weak moir\'e structure as clearly deduced in scanning tunnelling and atomic force microscopy (STM and AFM) experiments. These experiments combined with the respective density functional theory (DFT) results allow to discriminate between geometric and electronic contributions in the microscopy imaging of this system. Our spectroscopy (STS) and DFT results show that a graphene layer in the graphene/O/Ru(0001) system is electronically decoupled from the substrate and demonstrates the $p$-doping of the graphene-derived $\pi$ states. Analysis of the electronic structure of graphene, which can be accessed in ARPES experiments, is performed via unfolding of the graphene-related bands to the primitive $(1\times1)$ Brillouine zone of graphene.

\section*{Results and discussion}

\begin{figure}[t]
  \includegraphics[width=0.9\textwidth]{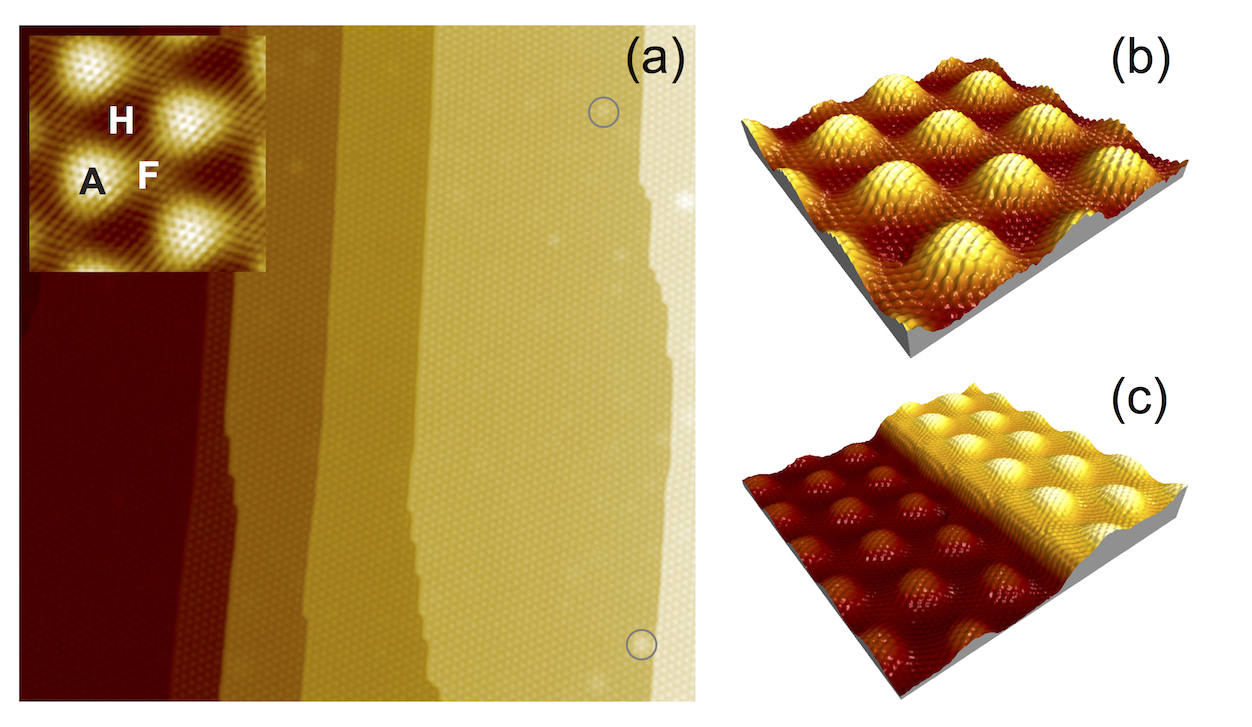}\\
  \caption{STM images of graphene/Ru(0001): (a) $230\times230\,\mathrm{nm}^2$, $U_T=0.3$\,V, $I_T=1.0$\,nA, (b) $9\times9\,\mathrm{nm}^2$, $U_T=0.15$\,V, $I_T=5.0$\,nA, (c) $14\times14\,\mathrm{nm}^2$, $U_T=0.15$\,V, $I_T=5.0$\,nA. Inset of (a) shows an atomically resolved zoomed area with all high-symmetry places of graphene/Ru(0001) marked by the respective capital letter, $6.1\times6.1\,\mathrm{nm}^2$, $U_T=-0.1$\,V, $I_T=5.0$\,nA.}
  \label{grRu0001_STM}
\end{figure}

Crystallographic and electronic structures of graphene/Ru(0001) and graphene/O/Ru(0001) were compared in local STM/STS and AFM experiments. Figure~\ref{grRu0001_STM} shows a large and small scale STM images of a complete graphene layer on Ru(0001). This system was prepared via decomposition of ethylene gas at $1300$\,K (for details, see Sec.~Methods). Such preparation leads to a graphene layer of very high quality with extremely small number of defects, which are mainly Ar atoms buried under interface Ru layer and they are marked by circles in Fig.~\ref{grRu0001_STM}(a). Hight quality of graphene is also confirmed by the atomically resolved STM images of the strongly corrugated graphene/Ru(0001) system where all high-symmetry positions of this system are clearly resolved: ATOP, HCP, and FCC [Fig.~\ref{grRu0001_STM}(a,b)]. They are marked by the corresponding capital letter in the inset of Fig.~\ref{grRu0001_STM}(a). Graphene covers Ru steps in a carpet fashion demonstrating the mirroring of the ATOP positions on the adjacent monoatomic steps in STM images of graphene/Ru(0001) [Fig.~\ref{grRu0001_STM}(c)]. This effect is assigned to the change of the atom stacking of the $hcp$ Ru(0001) substrate. Relatively strong interaction between graphene and Ru(0001) and the orbital intermixing of the graphene $\pi$ and Ru\,$4d$ states at the graphene/Ru interface makes it very difficult to distinguish separate carbon rings or atoms at step edges. This important fact will be later discussed for the graphene/O/Ru(0001) system where graphene is electronically decoupled from the substrate.

\begin{figure}
  \includegraphics[width=0.9\textwidth]{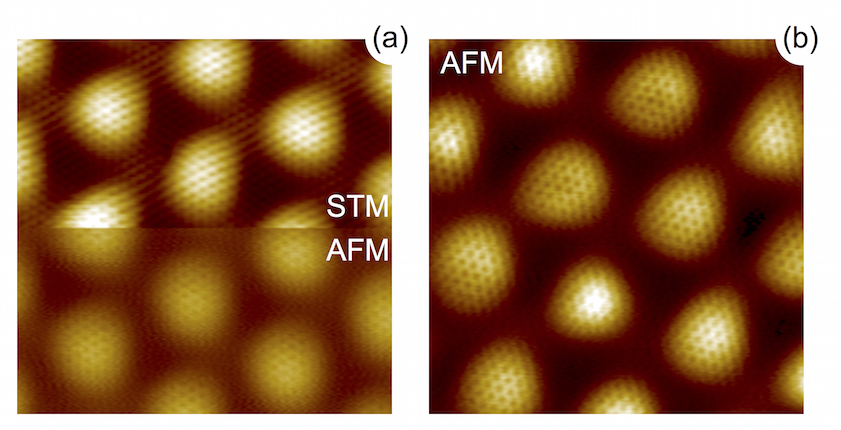}\\
  \caption{(a) Combined STM-top/AFM-bottom measurements of graphene/Ru(0001). Here the scanning mode was changed \textit{on-the-fly} in the middle of the scan area. Scanning parameters: $9.3\times9.3\,\mathrm{nm}^2$, $U_T=+0.05$\,V, $I_T=2.0$\,nA, $\Delta f=-3.7$\,Hz. (b) AFM image of graphene/Ru(0001). Scanning parameters: $8.7\times8.7\,\mathrm{nm}^2$, $U_T=+0.05$\,V, $\Delta f=-3.8$\,Hz.}
  \label{grRu0001_STM-AFM}
\end{figure}

Relatively large real space corrugation of $1.27$\,\AA\ of graphene on Ru(0001) (as deduced  from DFT calculations) makes AFM measurements with atomic resolution quite demanding. The results of such experiments are compiled in Fig.~\ref{grRu0001_STM-AFM} where combined STM/AFM data are shown. These figures show the raw experimental data and the possible image distortions are due to the relatively slow scanning velocity, which was adapted for the correct AFM experiments. In (a) the scanning mode was changed \textit{on-the-fly} in the middle of the scan that allows to carefully trace the imaging contrast between two modes and its relation to the geometric and electronic contributions. In our STM experiments for the bias voltages in the range of $\pm1$\,V, graphene/Ru(0001) is always imaged in the $direct$ imaging contrast while the highest ATOP places are imaged as bright areas and the low FCC and HCP areas as darker places. Moreover, no image contrast inversion is observed between STM and AFM contrary to graphene/Ir(111)~\cite{Voloshina:2013dq,Dedkov:2015iza}. Comparing these combined data with those from Fig.~\ref{grRu0001_STM} we can see that the imaging contrast and the relative areas for the high-symmetry places in STM depend on the tunnelling conditions ($U_T$ and $I_T$), but in AFM measurements the result is very close to the real geometry of the graphene/Ru(0001) system (relative areas of all places and the graphene corrugation). Strong corrugation of graphene in the present system gives a possibility to obtain clear atomic resolution in AFM only for the ATOP positions with a faint atomic contrast for the HCP and FCC places [Fig.~\ref{grRu0001_STM-AFM}(b)].

The oxygen intercalation was performed for the full graphene monolayer on Ru(0001) at $200^\circ$\,C and a partial pressure of $1.8\times10^{-4}$\,mbar of oxygen measured by the ion-gauge placed in the UHV chamber. Oxygen was introduced via a stainless steel pipe, which end was placed $1$\,mm from the sample surface that allows to increase the local pressure drastically (approximately by two orders of magnitude). This preparation procedure was initially also verified by the independent XPS measurements confirming our STM/AFM findings for the oxygen-intercalation system, graphene/O/Ru(0001).

\begin{figure}
  \includegraphics[width=0.9\textwidth]{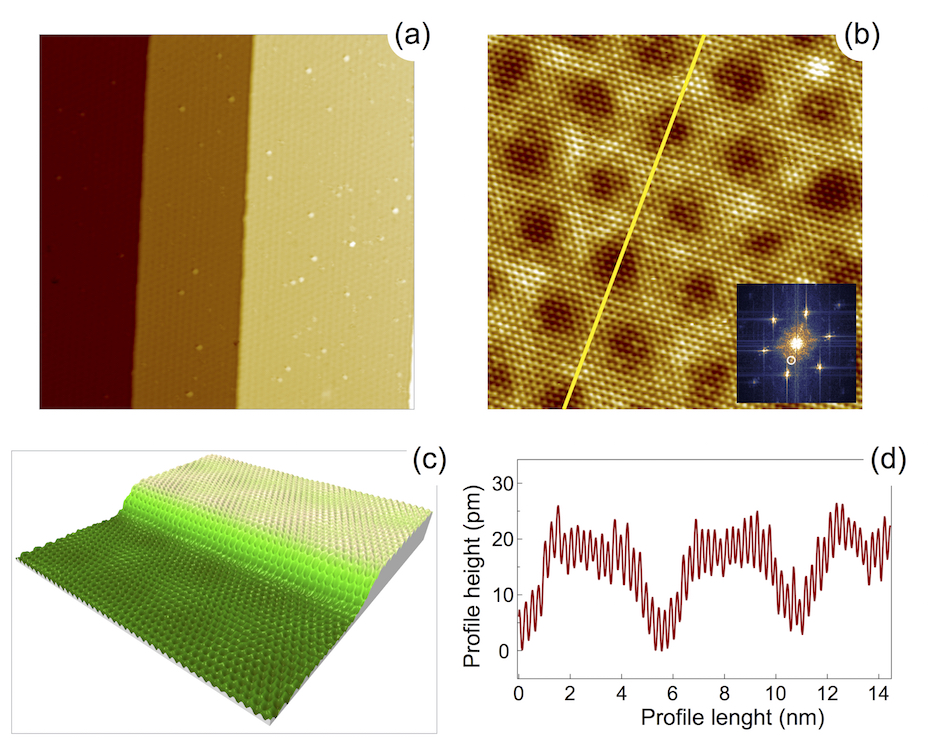}\\
  \caption{STM images of the graphene/O/Ru(0001) system: (a) $100\times100\,\mathrm{nm}^2$, $U_T=+0.3$\,V, $I_T=1.0$\,nA, (b) $13.5\times13.5\,\mathrm{nm}^2$, $U_T=+0.3$\,V, $I_T=2.0$\,nA. Inset shows the corresponding FFT image. (c) 3D rendering of the STM data of the region around step of graphene/O/Ru(0001). Scanning parameters: $8\times8\,\mathrm{nm}^2$, $U_T=+0.05$\,V, $I_T=1.0$\,nA. (d) Height profile of graphene/O/Ru(0001) marked by the yellow line in (b).}
  \label{grOxRu0001_STM}
\end{figure}

As can be found from the large scale STM image of graphene/O/Ru(0001) [Fig.~\ref{grOxRu0001_STM}(a)], oxygen fully intercalates in all sample areas. Resulting system presents the completely decoupled graphene layer from Ru substrate without any visible patches of the former graphene/Ru(0001) system. Similar to gr/Ru(0001) graphene forms a carpet which covers mono- and double-atomic steps of the gr/O/Ru(0001) system. Both, large and small scale, STM images of gr/O/Ru(0001) [Fig.~\ref{grOxRu0001_STM}(a,b)] demonstrate the weak moir\'e structure with the periodicity originating from the parent gr/Ru(0001) structure. The observed STM corrugation of graphene/O/Ru(0001) is smaller compared to the one of graphene/Ru(00001) and it depends on the imaging parameters, bias voltage ($U_T$) and tunneling current ($I_T$). For $U_T=+0.3$\,V/$I_T=2.0$\,nA it is $\approx20$\,pm [Fig.~\ref{grOxRu0001_STM}(b,d)] compared to $\approx95$\,pm measured at the same tunneling conditions for graphene/Ru(0001). This reduction of the graphene corrugation is a sign of its electronic decoupling from the Ru(0001) substrate.

Decoupling of graphene from the substrate and its nearly free-standing behaviour can be also concluded from the STM imaging of graphene/O/Ru(0001) around the step edges [Fig.~\ref{grOxRu0001_STM}(c)]. One can clearly see that contrary to graphene/Ru(0001), where strong modulation of the graphene structure along the step edge is found without clearly resolved carbon rings [Fig.~\ref{grRu0001_STM}(c)], for graphene/O/Ru(0001) only slight height variation of a graphene layer induced by moir\'e structure of the system is visible with clearly resolved carbon rings which flow over step edges. 

\begin{figure}
  \includegraphics[width=0.9\textwidth]{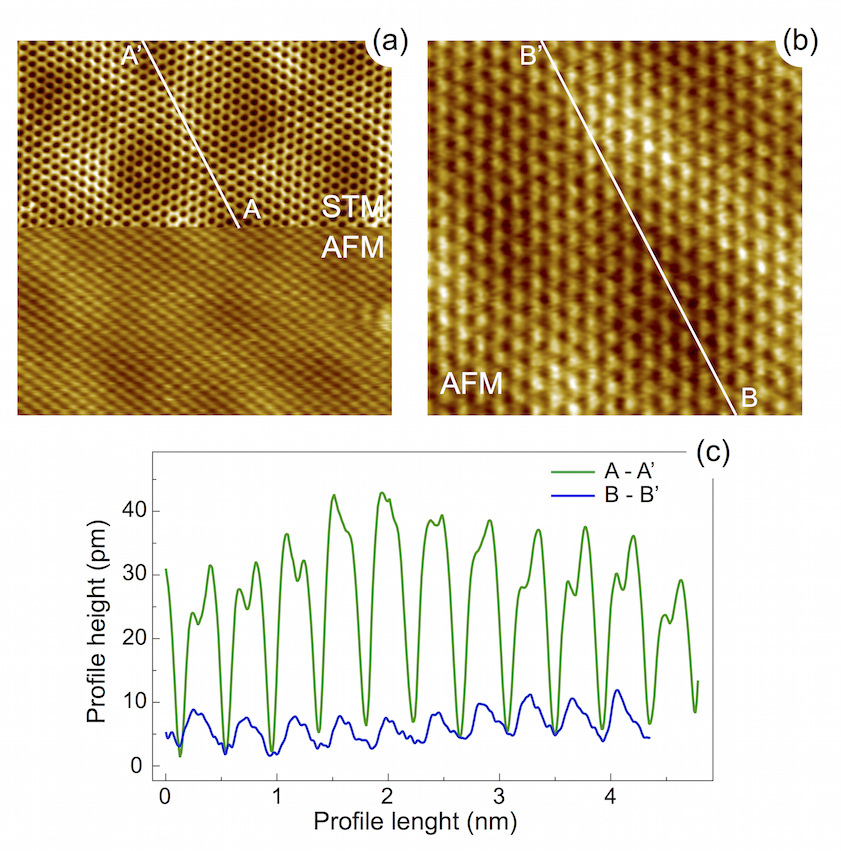}\\
  \caption{(a) Combined STM-top/AFM-bottom measurements of graphene/O/Ru(0001). Here the scanning mode was changed \textit{on-the-fly} in the middle of the scan area. Scanning parameters: $8.5\times8.5\,\mathrm{nm}^2$, $U_T=+0.05$\,V, $I_T=500$\,pA, $\Delta f=-3.9$\,Hz. (b) AFM image of graphene/O/Ru(0001). Scanning parameters: $4.1\times4.1\,\mathrm{nm}^2$, $U_T=+0.05$\,V, $\Delta f=-3.9$\,Hz. (c) Height profiles A-A' and B-B', extracted from (a) and (b), respectively.}
  \label{grOxRu0001_STM-AFM}
\end{figure}

Intercalated oxygen layer underneath graphene forms an ordered $(2\times1)$ structure on Ru(0001) as can be deduced from the Fast-Fourier-Transformation (FFT) image of the STM data shown in Fig.~\ref{grOxRu0001_STM}(b). Weak half-order spots might be taken as an indication for such structure (one of these spots is marked by the white circle; random adsorption of oxygen atoms might lead to the short order only without formation of big domains and inclusion in the analysis of the domain structure for the three-fold surface leads to the appearance of the the additional weak $(2\times2)$ structure in FFT). Here oxygen atoms form a saturation phase of $0.5$\,ML coverage on Ru(0001). This model was also confirmed by the recent micro-LEED (low energy electron diffraction) studies of this system~\cite{Sutter:2013kw}.

Similar to gr/Ru(0001) we also employ the combined STM/AFM measurements to the gr/O/Ru(0001) system. This allows to trace the true crystallographic structure of the system via discrimination between electronic and structural contributions in STM and AFM imaging. Figure~\ref{grOxRu0001_STM-AFM} shows the results of such experiments. Similar to previously shown results for gr/Ru(0001) the scanning mode in Fig.~\ref{grOxRu0001_STM-AFM}(a) was changed \textit{on-the-fly} in the middle of the scan. Both parts of this panel demonstrate the atomic resolution. Panel (b) of this figure shows the AFM scan of gr/O/Ru(0001) and the respective height profiles extracted from the similar places of STM and AFM are shown in panel (c). The extracted atomic and mean moir\'e lattice corrugations extracted from the STM data are $\approx30$\,pm and $\approx10$\,pm, respectively. The same values extracted from AFM images are $\approx6$\,pm and $\approx4$\,pm, respectively.    

After STM/AFM studies of gr/O/Ru(0001) the same sample was transferred through air (maximum exposure time was $1$\,min) in the other experimental station for low temperature STM/STS studies. Due to the protection properties of graphene and its extreme inertness~\cite{Dedkov:2008d,Dedkov:2008e,Borca:2009,Sutter:2010bx}, the studied system survives this transferring process and the clean surface of this system was recovered via the low temperature annealing $T=150^\circ$\,C in UHV conditions ($p<5\times10^{-10}$\,mbar). As was shown earlier by XPS~\cite{Dong:2015ig} such treatment does not lead to any change (intercalated oxygen desorption from the system) in the chemical composition of the gr/O/Ru(0001) system.

\begin{figure}
  \includegraphics[width=0.9\textwidth]{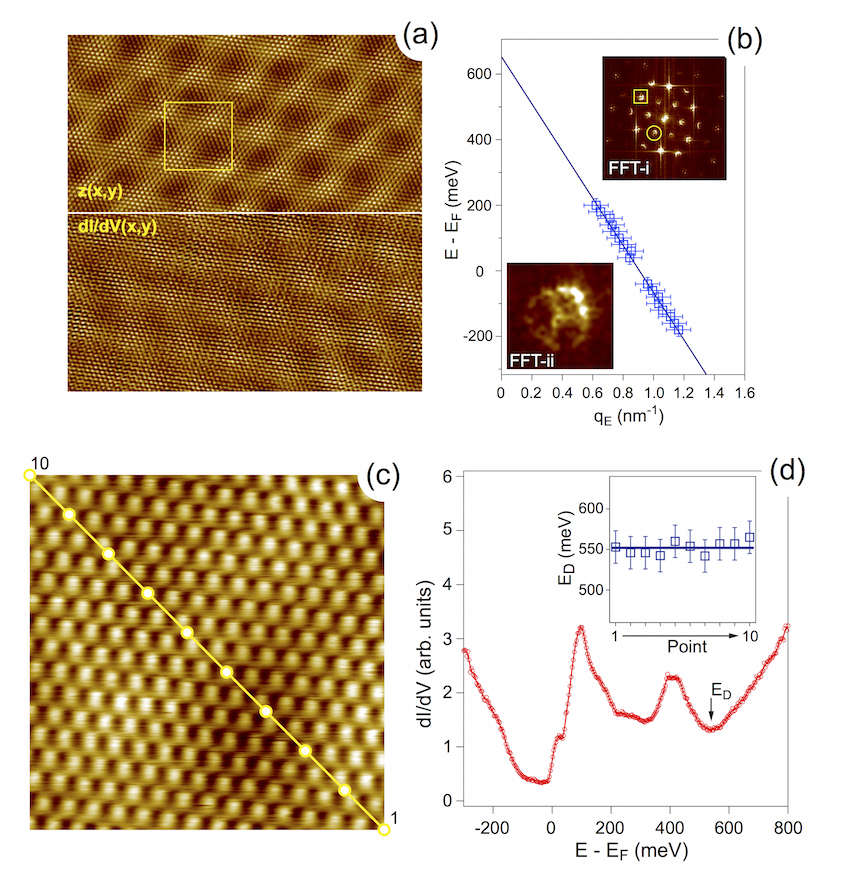}\\
  \caption{(a) Topography, $z(x,y)$, (top) and spectroscopy, $dI/dV(x,y)$, (bottom) maps of the gr/O/Ru(0001) system. Scanning parameters: $20\times 10 \mathrm{nm}^2$ (original scan size is $30\times 30\mathrm{nm}^2$), $U_T=+100$\,mV, $I_T=200$\,pA, $f_{mod}=684.74$\,Hz, $U_{mod}=10$\,mV. (b) Energy dispersion, $E(q_E)$, of the graphene-derived state around $E_F$ (open rectangles) together with the respective linear fit (solid line). Upper and lower insets show FFT images of the full $dI/dV$ map from (a). Lower inset is the zoom of the FFT image marked by the circle in the upper inset image. (c) Atomically resolved image of gr/O/Ru(0001) with the marks where single $dI/dV$ spectra were measured around $E_F$ (d). Scanning and spectroscopy parameters: $3.4\times 3.4\mathrm{nm}^2$, $U_T=+200$\,mV, $I_T=500$\,pA, $f_{mod}=684.74$\,Hz, $U_{mod}=10$\,mV. Inset of (d) shows the variation of the energy position of $E_D$ along the line in (c).}
  \label{grOxRu0001_dIdV}
\end{figure}

Figure~\ref{grOxRu0001_dIdV} shows a compilation of low temperature ($T=1$\,K) STM/STS studies of the graphene/ O/Ru(0001) system. Panel (a) presents combined topographic [$z(x,y)$] and STS [$dI/dV(x,y)$] maps (cut from the $30\times30\mathrm{nm}^2$) acquired simultaneously on this system at the bias voltage of $U_T=+100$\,mV. Both images demonstrate signal modulations on the moir\'e lattice and atomic scale. Atomic resolution and high statistic of the $dI/dV$ data allow to perform its careful FFT analysis and such result is shown as an upper inset in the panel (b) (FFT-i). On such 2D FFT image one can clearly recognise two kinds of features. The first one (one of these features is marked by the yellow rectangle) originates from the reciprocal lattice of graphene and it is also surrounded by 6 weak spots originating from the moir\'e lattice of the system. The second kind of the spectral features in the FFT map (one of them is marked by the yellow circle) reflects the intervalley scattering of the carriers in graphene around the $K$ points of the Brillouine zone (BZ) of graphene~\cite{Rutter:2007ep,Mallet:2012ib,Leicht:2014jy}. Such effect of scattering between two equivalent points in BZ, $K$ and $K'$, leads to the ring-like features in the FFT map of gr/O/Ru(0001) forming a $(\sqrt{3}\times\sqrt{3})R30^\circ$ structure with respect to the reciprocal lattice of graphene. The radius of these rings is $2q_E$, where $q_E$ is the wave vector of the Dirac particles at energy $E$ relative to $E_F$ and it is measured with respect to the $K$ point of the graphene BZ. [The possible deviation of the shape of the later features from the circle (see lower inset (FFT-ii) in the panel (b) might be due to the trigonal warping of the scattering vector map at higher binding energies with respect to $E_D$ as was shown in theoretical calculations~\cite{Simon:2011dv}. In this case electrons scattered at particular energy might have different wave vectors along $\Gamma-\mathrm{K}$ and $\mathrm{M}-\mathrm{K}$ directions around the $\mathrm{K}$-point. However in our analysis we ignore such distortions and approximate the intensity contour with circle because such simplification does not influence dramatically the extracted position of $E_D$ and Fermi velocity of carriers, $v_F$.]

Series of the topographic [$z(x,y)$] and spectroscopy [$dI/dV(x,y)$] maps measured at different bias voltages allow to get a series of the FFT maps and to extract the wave vector value of carriers, thus to obtain their energy dispersion, $E(q_E)$, around $E_F$. Such dispersion is shown in Fig.~\ref{grOxRu0001_dIdV}(b) as an array of blue squares. The respective fit of this data with the linear dependence, $E(q_E)=\hbar v_F q_E + E_D$ gives $v_F=(1.06\pm0.04)\cdot10^6$\,m/s and $E_D=610\pm50$\,meV above $E_F$, i.\,e. graphene layer is strongly $p$-doped.

In order to verify this result the single $dI/dV$ curves as a function of the bias voltage around $E_F$ were measured along the moir\'e lattice of gr/O/Ru(0001) as shown in Fig.~\ref{grOxRu0001_dIdV}(c). The representative $dI/dV$ curve (measured at the 3rd point of the series) is shown in Fig.~\ref{grOxRu0001_dIdV}(d) with the marked position of $E_D$ in the spectrum, that was identified as a local minima around the value obtained earlier from the $dI/dV$ maps. The respective variation of the energy position of this minima along the line from (c) is shown as an inset in panel (d). As one can see, these variations are within the error bar around the mean value for the position of the Dirac point at $552\pm20$\,meV and this value is also in rather good agreement with the previously obtained value deduced from a linear fit of the dispersion points around $E_F$ for the carriers in the gr/O/Ru(0001) system [Fig.~\ref{grOxRu0001_dIdV}(c)]. Our values for the $E_D$ energy position correlates with previously estimated position from ARPES data ($\sim0.5$\,eV above $E_F$)~\cite{Sutter:2010a}. However,  previous $dI/dV$ measurements reports $n$-doping of graphene in gr/O/Ru(0001) with $E_D=-0.48$\,eV~\cite{Jang:2013cn}, that can be considered as an artefact or misinterpretation of the experimental data.

\begin{figure}
  \includegraphics[width=\textwidth]{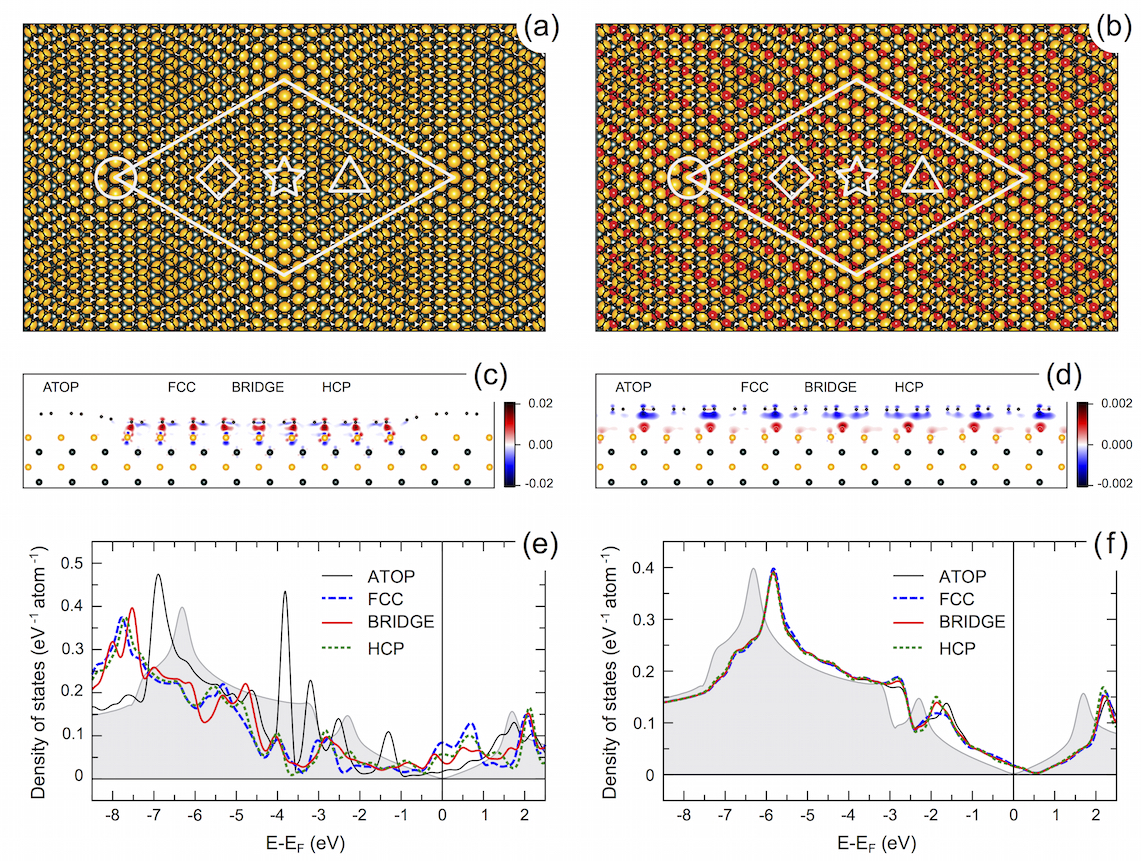}\\
  \caption{Top and side view of (a,c) graphene/Ru(0001) and (b,d) graphene/0.5 ML-O-($2\times1$)/Ru(0001) obtained after the structural optimization. The high-symmetry places in (a) and (b) are marked by circle, rhombus, star, and triangle for ATOP, FCC, BRIDGE, and HCP positions, respectively. In (b) and (d) the structure of the respective system is overlaid with the calculated difference electron density, $\Delta \rho_\textrm{gr/sub}(r)= \rho_\textrm{gr/sub}(r)-(\rho_\textrm{gr}(r)+\rho_\textrm{sub}(r))$ [sub = substrate, \textit{i.\,e.} either Ru(0001) or 0.5 ML-O-($2\times1$)/Ru(0001)], plotted in units of $e/\textrm{\AA}^3$. Carbon-site projected density of states calculated for all high-symmetry places of (e) graphene/Ru(0001) and (f) graphene/0.5 ML-O-($2\times1$)/Ru(0001). Greyed plot in (e) and (f) shows the corresponding DOS for the free-standing graphene. }
\label{grRu-grORu_StrDOS}
\end{figure}

In order to justify our experimental findings we perform analysis of these data within the framework of DFT (for computational details, see Sec.~Methods). Figure~\ref{grRu-grORu_StrDOS} shows (a and b) top and (c and d) side views of the gr/Ru(0001) and gr/$0.5$\,ML-O-$(2\times1)$/Ru(0001) structures, respectively, obtained with optimized geometries. (The corresponding results for the gr/$1$\,ML-O-$(1\times1)$/Ru(0001) system are presented in Supplementary materials: Fig.\,S1). In panels (c) and (d) the side view is overlaid with the corresponding difference electron density, $\Delta\rho(r)$, at the interface. According to the presented results, graphene on Ru(0001) is a strongly corrugated system ($\Delta h=1.27$\,\AA, see Supplementary materials: Fig.\,S2 (a)) consisted of the regions of the alternating \textit{weak} and \textit{strong} interaction between graphene and Ru substrate with the minimal distance between C and Ru of $2.24$\,\AA\ (that corresponds to a minimal interlayer distance between graphene and Ru(0001) of $2.10$\,\AA) [Fig.~\ref{grRu-grORu_StrDOS}(c)]. The original electronic structure of free-standing graphene is strongly modified after its adsorption on Ru due to the charge transfer from Ru to C and strong orbital overlap of the valence band states of graphene and Ru at the interface [Fig.~\ref{grRu-grORu_StrDOS}(e)]. This can be indicated by the formation of the \textit{covalent-like} bonds at the graphene-Ru interface for the \textit{strongly} interacting places of the structure. These results are in very good agreement with previous works~\cite{Wang:2008,Wang:2010jw,Stradi:2011be,Stradi:2013dj}.

According to the previous~\cite{Sutter:2010a,Sutter:2013kw} and present findings (Fig.~\ref{grOxRu0001_STM}) oxygen forms a stable $0.5$\,ML-O-$(2\times1)$ phase on Ru(0001). Therefore the oxygen-intercalated system in our studies was modelled as gr/$0.5$\,ML-O-$(2\times1)$/Ru(0001) where oxygen atoms are placed in the $hcp$ adsorption sites of Ru(0001). The resulting stable structure of this system (top and side views) is shown in Fig.~\ref{grRu-grORu_StrDOS}(b,d). One can see that a graphene layer here is almost flat with a corrugation of $0.12$\,\AA, which is very close to the value found in our AFM experiments [Fig.~\ref{grOxRu0001_STM-AFM}(b,c)]. The mean distance between graphene and oxygen is $2.78$\,\AA, which is close to the typical values for the intercalation graphene $p$-doped systems.

The side view of gr/$0.5$\,ML-O-$(2\times1)$/Ru(0001) is overlaid with the difference charge density [Fig.~\ref{grRu-grORu_StrDOS}(d)], which shows the decoupling of graphene from the substrate. There is no orbital overlap of the valence band state of graphene and substrate, and charge depletion (accumulation) takes place on graphene (O-Ru) that leads to $p$-doping of graphene and formation of the localized dipole moment at the graphene-O-Ru interface.

The electronic structure of graphene (DOS) on Ru(0001) and on $0.5$\,ML-O-$(2\times1)$/Ru(0001) is shown in Fig.~\ref{grRu-grORu_StrDOS} (e) and (f), respectively. In the former case, the orbital overlap of the valence band states of graphene and Ru leads to the strong modification of the electronic structure of a graphene layer (see for comparison DOS for free-standing graphene). Site-selective interaction in gr/Ru(0001) leads to different site-projected C-DOS as can be seen in Fig.~\ref{grRu-grORu_StrDOS}(e) and several localized so-called \textit{interface states} are formed in the energy range $E-E_F=-2 \ldots +2$\,eV (see Supplementary material: Fig.\,S3). The doping level of graphene in this case can be estimated from the position of minima in DOS curve for the ATOP high-symmetry position: $E-E_F=-902$\,meV, i.\,e. as expected graphene on Ru(0001) is strongly $n$-doped.

Oxygen intercalation ($0.5$\,ML) in graphene/Ru(0001) leads to the almost complete recovering of the free-standing character of the graphene-derived electronic states. Comparing DOS for this system with the one for free-standing graphene, the position of Dirac point is $E_D=+536$\,meV, which is in very good agreement with the values obtained in our STM/STS experiments. For the gr/$1$\,ML-O-$(1\times1)$/Ru(0001) system the doping is higher [$E_D=+638$\,meV, see Supplementary material: Fig.\,S1 (c)]. The small difference in the site-projected C-DOS for gr/$0.5$\,ML-O-$(2\times1)$/Ru(0001) at $E-E_F=-3 \ldots -1$\,eV is in the energy range corresponding to the van Hove singularities for the $\sigma$ and $\pi$ states at the $\Gamma$ and $\mathrm{M}$ points of BZ, respectively, and therefore their energy positions and intensities might be affected, for such big system (total number of atoms is $986$), by the slight variation of the $k$-points sampling in BZ during integration. 

\begin{figure}[t]
  \includegraphics[width=1.0\textwidth]{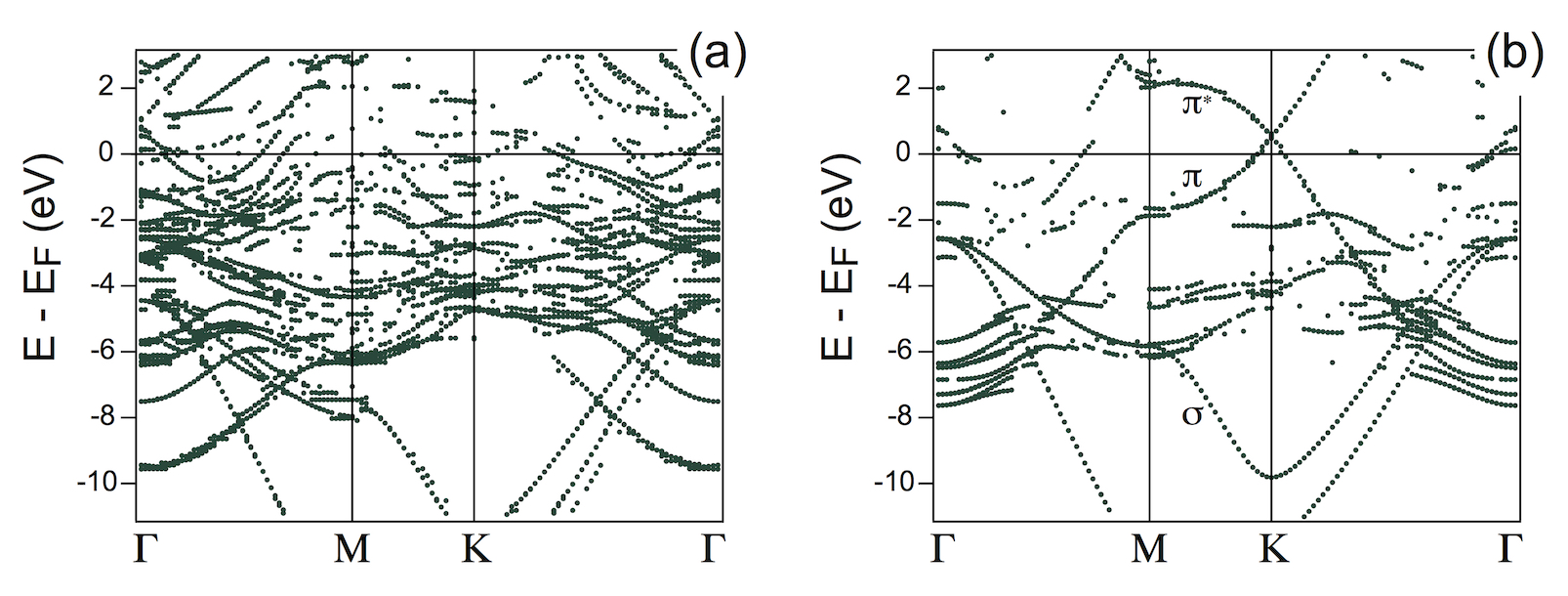}\\
  \caption{Band structure of (a) graphene/Ru(0001) and (b) graphene/$0.5$\,ML-O-$(2\times1)$/Ru(0001) unfolded for the graphene ($1\times1$) primitive cell. }
  \label{grRu-grORu_bands}
\end{figure}

Similar conclusion regarding effective electronic decoupling of graphene from metal via oxygen intercalation can be made when comparing the results of band structure calculations performed for the studied systems [Fig.~\ref{grRu-grORu_bands} (a,b)]. For graphene/Ru(0001) one can clearly see the result typical for the \textit{strongly} interacting system with the formation of several interface states at the K-point of BZ as a result of hybridisation between graphene and valence band states leading to the complete destruction of the Dirac cone. Thus, as it was already postulated in Ref.~\cite{Voloshina:2012c}, in spite of the coexistence of the \textit{weakly} and \textit{strongly} interacting regions within the graphene/Ru(0001) system, due to the metallic character of graphene in this system, the electronic structure of graphene in its original BZ is defined by the bonding strength and the electronic structure at the most perturbed graphene places. Upon the oxygen intercalation hybridisation between graphene and metal valence band states is absent (see Supplementary material: Fig.\,S4). The main $\pi$ (as well as $\sigma$) branches are clearly recognisable in the electronic structure of graphene/$0.5$\,ML-O-$(2\times1)$/Ru(0001) [Fig.~\ref{grRu-grORu_bands} (b)] and they almost reproduce the electronic structure of the free-standing graphene (except for a upwards shift due to the graphene doping). It is interesting to note the absence of the gap at the Dirac point that is in good agreement with the model describing the electronic structure of graphene modified by its interaction with a substrate~\cite{Voloshina:2014jl}.

\begin{figure}[t]
  \includegraphics[width=\textwidth]{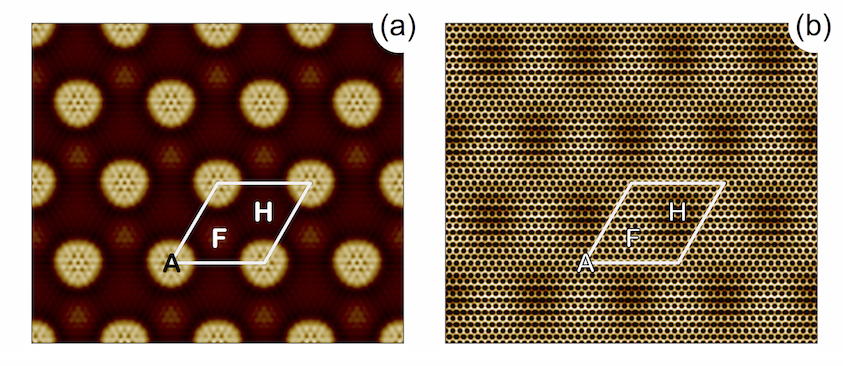}\\
  \caption{Calculated CC STM images of (a) graphene/Ru(0001) and (b) graphene/0.5 ML-O-($2\times1$)/Ru(0001) obtained at $+300$\,mV of the bias voltage. The high-symmetry places are marked by the respective capital letters.}
  \label{grRu-grORu_ThSTM}
\end{figure}

In order to prove our results we also perform simulations of STM images in the framework of the Tersoff-Hamann formalism~\cite{Tersoff:1985}. Figure~\ref{grRu-grORu_ThSTM} shows constant-current STM images of (a) gr/Ru(0001) and (b) gr/$0.5$\,ML-O-$(2\times1)$/Ru(0001). The presented results are in very good agreement with experimental images [Figs.~\ref{grOxRu0001_STM}(b), \ref{grOxRu0001_STM-AFM}(a), \ref{grOxRu0001_dIdV}(a)]. The moir\'e unit cells of gr/O/Ru(0001) in the experimental STM images are slightly elongated due to the respective structure of $0.5$\,ML-O-$(2\times1)$/Ru(0001) underneath graphene. For the gr/$1$\,ML-O-$(1\times1)$/Ru(0001) system the STM images do not show such an effect (see Supplementary material: Fig.\,S5).

\section*{Conclusion}

In our work we successfully decouple a \textit{strongly} interacting graphene layer from the Ru(0001) substrate via intercalation of an oxygen layer. The modifications of the crystallographic structure before and after intercalation are traced by the combined STM/AFM measurements. Strong variation of the imaging contrast is found in STM, reflecting the electronic structure of the system, whereas the corrugation obtained in AFM experiments is very close to the one deduced from DFT calculations. Decoupling of graphene via oxygen intercalation leads to the strong $p$-doping of a graphene layer with the Dirac point located at $E-E_F=+552$\,meV above $E_F$ as found in both STS and $dI/dV(U_T)$-mapping experiments. This value is in very good agreement with the doping level obtained in DFT calculations. We employ the graphene-derived bands unfolding procedure in order to obtain the dispersion of the electronic states $E(k)$ with respect to the $(1\times1)$ BZ of graphene. This allows us to directly trace the band structure in the vicinity of $E_F$ and $E_D$ and compare this results with our STS data and future ARPES experiments.

\section*{Methods}

\textbf{Sample Preparation.} Graphene/Ru(0001) system was prepared in ultrahigh vacuum system via cracking of ethylene: $T=1020$\,K, $p=2\times10^{-7}$mbar, $t=30$\,min. This procedure leads to the single-domain graphene layer on Ru(0001) of very high quality, which was verified by means of STM. Intercalation of oxygen in graphene/Ru(0001) was performed at $473$\,K for $30$\,min when oxygen was introduced in UHV chamber via a stainless steel pipe which end was placed about $1$\,mm from the sample surface (this allows to increase the local pressure drastically). The oxygen partial pressure during intercalation, measured by the ion-gauge, was $p=1.8\times10^{-4}$mbar (from our estimations the local pressure at the sample surface is higher by 2 orders of magnitude). The result of this intercalation process and quality of the gr/O/Ru(0001) was verified by the large scale STM measurements. 

\textbf{STM and AFM Experiments.} The STM and AFM measurements were performed in constant current or constant frequency shift modes, respectively. In this case the topography of sample, $z(x, y)$, is studied with the corresponding signal, tunnelling current $(I_T)$ or frequency shift $(\Delta f)$, used as an input for the feedback loop. The STM/AFM images were collected at room temperature with SPM Aarhus 150 equipped with KolibriSensor\texttrademark\ from SPECS with Nanonis Control system. In these measurements the sharp W-tip was used which was cleaned in situ via Ar$^+$-sputtering. In presented STM images the tunnelling bias voltage, $U_T$, is applied to the sample and the tunnelling current, $I_T$, is collected through the tip, which is virtually grounded. During the AFM measurements the sensor was oscillating with the resonance frequency of $f_0=998666$\,Hz and the quality factor of $Q=22500$. The oscillation amplitude was set to $A=300$\,pm. Low-temperature STM and $dI/dV$ measurements were performed in the SPECS JT-STM at the sample and tip temperature of $1$\,K. Tunneling current and voltage values are given in the figure captions. $dI/dV$ mappings were performed at low temperatures using the lock-in-technique with modulation voltage of $U_{mod}=10$\,mV and modulation frequency $f_{mod}=684.74$\,Hz. Base pressure in both experimental stations is below $5\times10^{-11}$\,mbar.

\textbf{DFT Calculations.}  The DFT calculations were carried out using the projector augmented wave (PAW) method~\cite{Blochl:1994}, a plane wave basis set and the generalized gradient approximation as parameterized by Perdew \textit{et al.}~\cite{Perdew:1996}, as implemented in the VASP program~\cite{Kresse:1994}. The plane wave kinetic energy cutoff was set to $400$\,eV. The long-range van der Waals interactions were accounted for by means of the DFT-D2 approach~\cite{Grimme:2006}. The corresponding structures of the graphene-metal based systems are shown in Fig.~\ref{grRu-grORu_StrDOS} (a,c) and they are discussed in details in the text. The supercell used to model the graphene-metal interface has a ($12\times12$) lateral periodicity with respect to Ru(0001). It is constructed from a slab of $4$ layers of Ru atoms with a ($13\times13$) graphene layer adsorbed from one side and a vacuum region of approximately $20$\,\AA. In the case of graphene/O/Ru(0001) systems an oxygen layer (either full or half) was inserted between graphene and Ru(0001). To avoid interactions between periodic images of the slab, a dipole correction is applied~\cite{}. During the structure relaxation, the positions of the carbon atoms as well as those of the top two layers of metal atoms (as well as oxygen atoms in the case of graphene/O/Ru(0001)) are allowed to relax. In the total energy calculations and during the structural relaxations the \textbf{k}-meshes for sampling the supercell Brillouin zone are chosen to be as dense as $6\times6$ and $3\times3$, respectively, and centred at the $\Gamma$-point. The STM images are calculated using the Tersoff-Hamann formalism~\cite{Tersoff:1985}. The band structures calculated for the studied systems were unfolded to the graphene ($1\times1$) primitive unit cell according to the procedure described in Refs.~\cite{Popescu:2012bq,Medeiros:2014ka} with the code BandUP~\cite{Medeiros:2014ka}.


\section*{Acknowledgements}

The authors thank Dr. Silvano Lizzit (Trieste, Italy) and Dr. Rosanna Larciprete (Rome, Italy) for useful discussions. The High Performance Computing Network of Northern Germany (HLRN-III) is acknowledged for computer time. Financial support from the German Research Foundation (DFG) through the grant VO1711/3-1 within the Priority Programme 1459 ``Graphene'' is appreciated. 

\section*{Author contributions}

E.V. performed DFT calculations. N.B. and Y.D. performed scanning probe microscopy experiments. All authors contribute in the analysis of data and writing of manuscript.

\section*{Additional information}

\textbf{Competing financial interests:} The authors declare no competing financial interests.

\newpage

\noindent
Supplementary material for manuscript:\\
\textbf{Restoring a free-standing character of graphene on Ru(0001)}\\
\newline
Elena Voloshina,$^1$ Nikolai Berdunov,$^2$ and Yuriy Dedkov$^{2}$\\
\newline
$^1$Humboldt-Universit\"at zu Berlin, Institut f\"ur Chemie, 10099 Berlin, Germany\\
$^2$SPECS Surface Nano Analysis GmbH, Voltastra\ss e 5, 13355 Berlin, Germany\\

\noindent\textbf{List of figures:}

\noindent\textbf{Fig.\,S1.} (a) Top and side view of the graphene/1ML O/Ru(0001) obtained after the structural optimization. The high-symmetry places in (a) are marked by circle, rhombus, star, and triangle for ATOP, FCC, BRIDGE, and HCP positions, respectively. In (b) the structure of the respective system is overlaid with the calculated difference electron density, $\Delta \rho_{\textrm{gr/O/Ru}}(r)= \rho_\textrm{gr/O/Ru}(r)-(\rho_\textrm{gr}(r)+\rho_\textrm{O/Ru}(r))$, plotted in units of $e/\textrm{\AA}^3$. (c) Carbon-site projected density of states calculated for all high-symmetry places of graphene/1ML O/Ru(0001). Greyed plot in (c) shows the corresponding DOS for the free-standing graphene.

\noindent\textbf{Fig.\,S2.} The height variation of the carbon atoms in the graphene layer on (a) Ru(0001), (b) 0.5 ML-O-($2\times1$)/Ru(0001), and (c) 1 ML-O/Ru(0001). The colour bar on the left-hand side represents the hight scale in \AA.

\noindent\textbf{Fig.\,S3.} (a) Carbon-projected DOSs for different high-symmetry places of the graphene/Ru(0001) system. (b) Ru-projected DOS for graphene/Ru(0001).

\noindent\textbf{Fig.\,S4.} (a) Carbon-projected DOSs for different high-symmetry places of the graphene/0.5 ML-O-($2\times1$)/Ru(0001) system. (b) Oxygen-projected DOS for graphene/0.5 ML-O-($2\times1$)/Ru(0001). (c) DOS projected onto Ru-atoms of the topmost metal-layer for graphene/0.5 ML-O-($2\times1$)/Ru(0001).

\noindent\textbf{Fig.\,S5.} Calculated CC STM images of (a) graphene/Ru(0001), (b) graphene/0.5 ML-O-($2\times1$)/Ru(0001), and (c) graphene/1 ML-O/Ru(0001) obtained at $300$\,meV of the bias voltage.

\newpage
\begin{figure}[h]
  \includegraphics[width=0.6\textwidth]{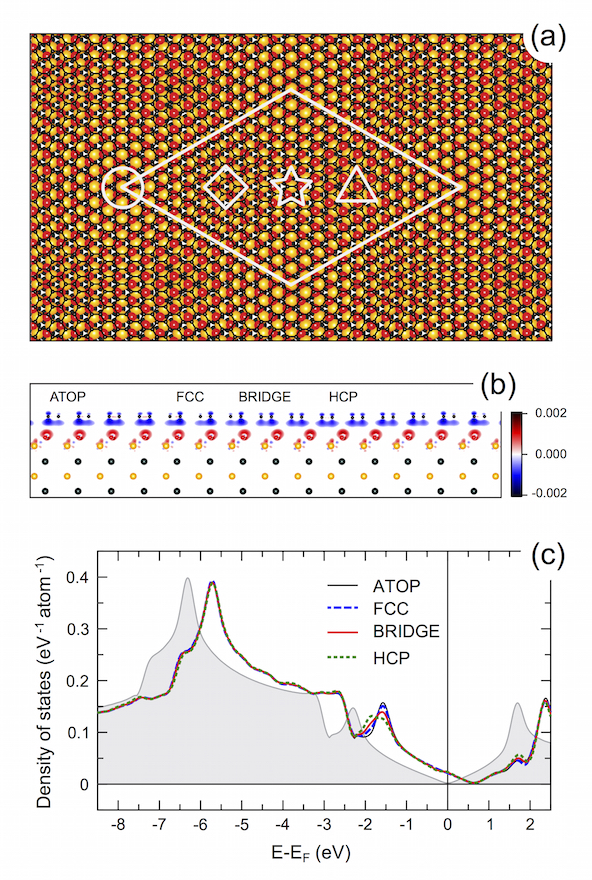}\\
 \label{grOxRu_StrDOS}
\end{figure}
\noindent\textbf{Fig.\,S1.} (a) Top and side view of the graphene/1ML O/Ru(0001) obtained after the structural optimization. The high-symmetry places in (a) are marked by circle, rhombus, star, and triangle for ATOP, FCC, BRIDGE, and HCP positions, respectively. In (b) the structure of the respective system is overlaid with the calculated difference electron density, $\Delta \rho_\textrm{gr/O/Ru}(r)= \rho_\textrm{gr/O/Ru}(r)-(\rho_\textrm{gr}(r)+\rho_\textrm{O/Ru}(r))$, plotted in units of $e/\textrm{\AA}^3$. (c) Carbon-site projected density of states calculated for all high-symmetry places of graphene/1ML O/Ru(0001). Greyed plot in (c) shows the corresponding DOS for the free-standing graphene.

\newpage
\begin{figure}[h]
  \includegraphics[width=0.5\textwidth]{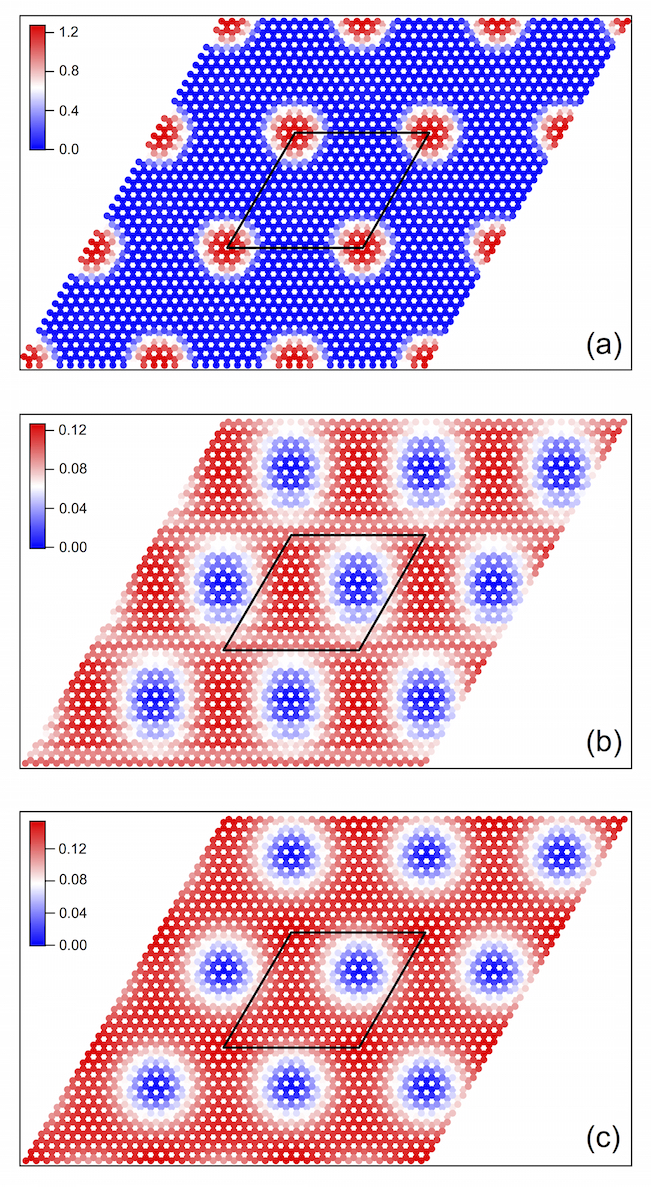}\\
   \label{C-height}
\end{figure}
\noindent\textbf{Fig.\,S2.} The height variation of the carbon atoms in the graphene layer on (a) Ru(0001), (b) 0.5 ML-O-($2\times1$)/Ru(0001), and (c) 1 ML-O/Ru(0001). The colour bar on the left-hand side represents the hight scale in \AA.

\newpage
\begin{figure}[h]
  \includegraphics[width=0.6\textwidth]{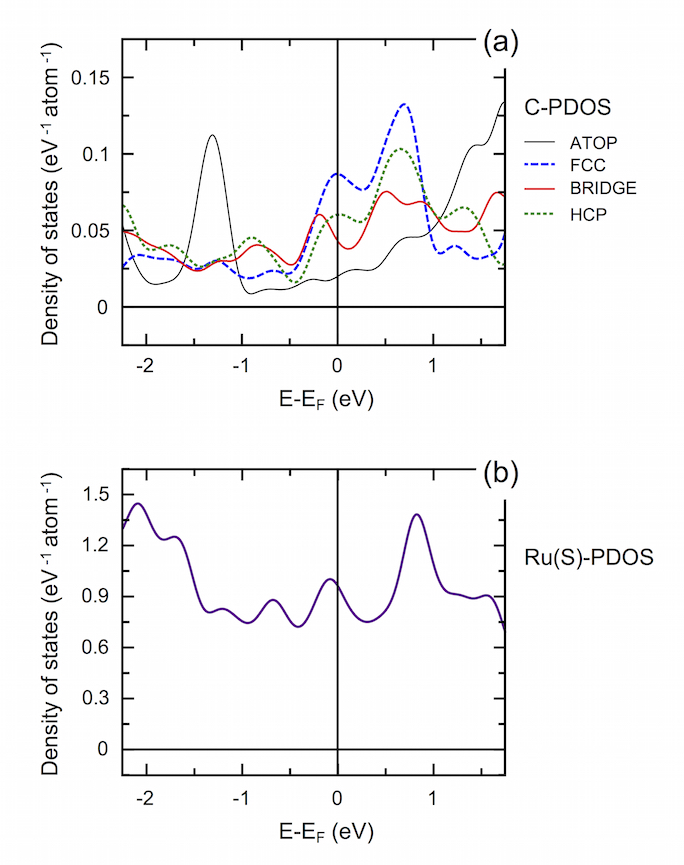}\\
   \label{grRu_PDOS}
  \end{figure}
\noindent\textbf{Fig.\,S3.} (a) Carbon-projected DOSs for different high-symmetry places of the graphene/Ru(0001) system. (b) Ru-projected DOS for graphene/Ru(0001).

\newpage
\begin{figure}[h]
  \includegraphics[width=0.6\textwidth]{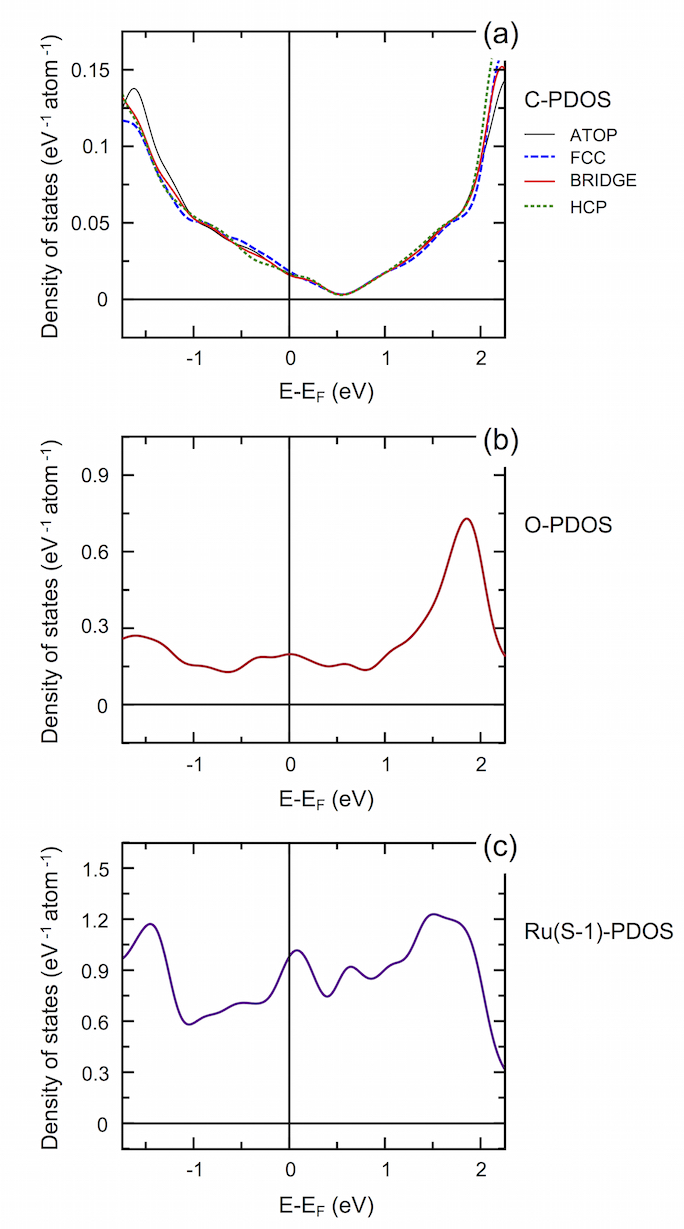}\\
     \label{grORu_PDOS}
\end{figure}
\noindent\textbf{Fig.\,S4.} (a) Carbon-projected DOSs for different high-symmetry places of the graphene/0.5 ML-O-($2\times1$)/Ru(0001) system. (b) Oxygen-projected DOS for graphene/0.5 ML-O-($2\times1$)/Ru(0001). (c) DOS projected onto Ru-atoms of the topmost metal-layer for graphene/0.5 ML-O-($2\times1$)/Ru(0001).

\newpage
\begin{figure}[h]
  \includegraphics[width=0.45\textwidth]{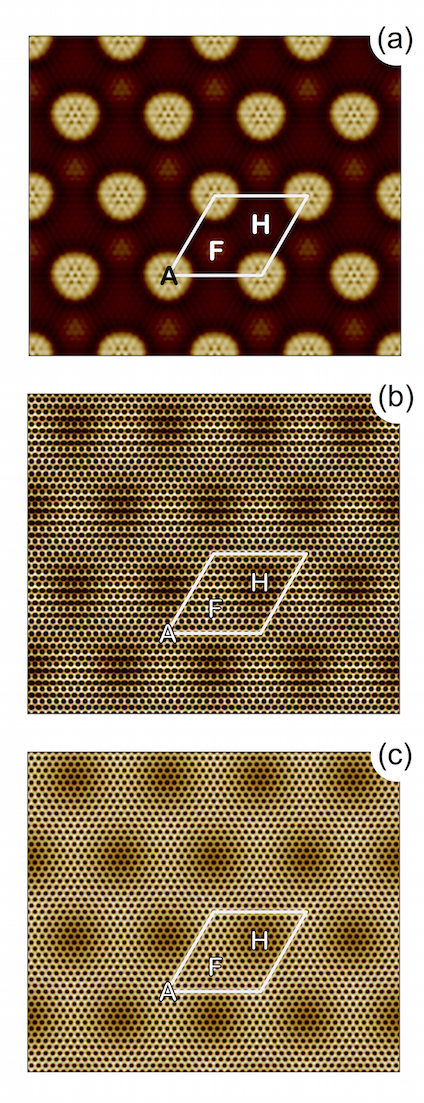}\\
  \label{stm}
\end{figure}
\noindent\textbf{Fig.\,S5.} Calculated CC STM images of (a) graphene/Ru(0001), (b) graphene/0.5 ML-O-($2\times1$)/Ru(0001), and (c) graphene/1 ML-O/Ru(0001) obtained at $300$\,meV of the bias voltage.


\begin{thebibliography}{99}
\bibitem{Tontegode:1991ts}
Tontegode,~A.~Y. Carbon on transition metal surfaces. \emph{Prog. Surf. Sci.} \textbf{38}, 201--429 (1991).

\bibitem{Wintterlin:2009}
Wintterlin,~J. \& Bocquet,~M.~L. Graphene on metal surfaces. \emph{Surf. Sci.} \textbf{603}, 1841--1852 (2009).

\bibitem{Dedkov:2012book}
Dedkov, Y. S., Horn, K., Preobrajenskij, A. \& Fonin, M. Epitaxial graphene on metals. \textit{Graphene Nanoelectronics} Raza, H. (ed.) 189-234 (Springer, Berlin, 2012).

\bibitem{Batzill:2012}
Batzill,~M. The Surface science of graphene: metal interfaces, CVD synthesis, nanoribbons, chemical modifications, and defects. \emph{Surf. Sci. Rep.} \textbf{67}, 83--115 (2012).

\bibitem{Dedkov:2015kp}
Dedkov,~Y. \& Voloshina,~E. Graphene growth and properties on metal substrates. \emph{J Phys.: Condens. Matter} \textbf{27}, 303002 (2015).

\bibitem{Voloshina:2012c}
Voloshina,~E. \& Dedkov,~Y. Graphene on metallic surfaces: problems and perspectives. \emph{Phys. Chem. Chem. Phys.} \textbf{14}, 13502 (2012).

\bibitem{Voloshina:2014jl}
Voloshina,~E.~N. \& Dedkov,~Y.~S. General approach to understanding the electronic structure of graphene on metals. \emph{Materials Research Express} \textbf{1}, 035603 (2014).

\bibitem{Rutkov:1995}
Rutkov,~E.~V., Tontegode,~A.~Y. \& Usufov,~M.~M. Evidence for a C$_{60}$ monolayer intercalated between a graphite monolayer and iridium. \emph{Phys. Rev. Lett.} \textbf{74}, 758--760 (1995).

\bibitem{Shikin:2000a}
Shikin,~A., Dedkov,~Y., Adamchuk,~V., Farias,~D. \& Rieder,~K. Formation of an intercalation-like system by intercalation of C$_{60}$ molecules underneath a graphite monolayer on Ni(111). \emph{Surf. Sci.} \textbf{452}, 1--8 (2000).

\bibitem{Granas:2013tl}
Gr{\aa}n{\"a}s,~E. \textit{et al.} CO intercalation of graphene on Ir(111) in the millibar regime. \emph{J. Phys. Chem. C} \textbf{117}, 16438--16447 (2013).

\bibitem{Jin:2014cl}
Jin,~L. \textit{et al.} Surface chemistry of CO on Ru(0001) under the confinement of graphene cover. \emph{J. Phys. Chem. C} \textbf{118}, 12391--12398 (2014.

\bibitem{Zhang:2009qqq}
Zhang,~H., Fu,~Q., Cui,~Y., Tan,~D. \& Bao,~X. Growth mechanism of graphene on Ru(0001) and O$_2$ adsorption on the graphene/Ru(0001) surface. \emph{J. Phys. Chem. C} \textbf{113}, 8296--8301 (2009).
  
 \bibitem{Sutter:2010a}
Sutter,~P., Sadowski,~J.~T. \& Sutter,~E.~A. Chemistry under cover: tuning metal-graphene interaction by reactive intercalation. \emph{J. Am. Chem. Soc.} \textbf{132}, 8175--8179 (2010).

\bibitem{Larciprete:2012aaa}
Larciprete,~R. \textit{et al.} Oxygen switching of the epitaxial graphene-metal interaction. \emph{ACS Nano} \textbf{6}, 9551--9558 (2012).

\bibitem{Granas:2012cf}
Gr{\aa}n{\"a}s,~E. \textit{et al.} Oxygen intercalation under graphene on Ir(111): energetics, kinetics, and the role of graphene edges. \emph{ACS Nano} \textbf{6}, 9951--9963 (2012).

\bibitem{Jin:2012ki}
Jin,~L. \textit{et al.} Tailoring the growth of graphene on Ru(0001) via engineering of the substrate surface. \emph{J. Phys. Chem. C} \textbf{116}, 2988--2993 (2012).

\bibitem{Sutter:2013kw}
Sutter,~P., Albrecht,~P., Tong,~X. \& Sutter,~E. Mechanical decoupling of graphene from Ru(0001) by interfacial reaction with oxygen. \emph{J. Phys. Chem. C} \textbf{17}, 6320--6324 (2013).

\bibitem{Jang:2013cn}
Jang,~W.~J., Kim,~H., Jeon,~J.~H., Yoon,~J.~K. \& Kahng,~S.-J. Recovery and local-variation of Dirac cones in oxygen-intercalated graphene on Ru(0001) studied using scanning tunneling microscopy and spectroscopy. \emph{Phys. Chem. Chem. Phys.} \textbf{15}, 16019--16023 (2013).

\bibitem{Dong:2015ig}
Dong,~A. \textit{et al.} Facile oxygen intercalation between full layer graphene and Ru(0001) under ambient conditions. \emph{Surf. Sci.} \textbf{634}, 37--43 (2015).

\bibitem{Voloshina:2013dq}
Voloshina,~E.~N. \textit{et al.} Electronic structure and imaging contrast of graphene moir\'e on metals. \emph{Sci. Rep.} \textbf{3}, 1072 (2013).

\bibitem{Dedkov:2015iza}
Dedkov,~Y., Voloshina,~E. \& Fonin,~M. Scanning probe microscopy and spectroscopy of graphene on metals. \emph{Phys. Status Solidi B} \textit{252}, 451--468 (2015).

\bibitem{Dedkov:2008d}
Dedkov,~Y.~S., Fonin,~M. \& Laubschat,~C. A possible source of spin-polarized electrons: The inert graphene/Ni(111) system. \emph{Appl. Phys. Lett.} \textbf{92}, 052506 (2008).

\bibitem{Dedkov:2008e}
Dedkov,~Y.~S., Fonin,~M., Ruediger,~U. \& Laubschat,~C. Graphene-protected iron layer on Ni(111). \emph{Appl. Phys. Lett.}
\textbf{93}, 022509 (2008).

\bibitem{Borca:2009}
Borca,~B., Calleja,~F., Hinarejos,~J.~J., Vazquez~de Parga,~A.~L. \& Miranda,~R. Reactivity of periodically rippled graphene grown on Ru(0001).
  \emph{J Phys.: Condens. Matter} \textbf{21}, 134002 (2009).

\bibitem{Sutter:2010bx}
Sutter,~E., Albrecht,~P., Camino,~F.~E. \& Sutter,~P. Monolayer graphene as ultimate chemical passivation layer for arbitrarily shaped metal surfaces. \emph{Carbon} \textbf{48}, 4414 (2010).

\bibitem{Rutter:2007ep}
Rutter,~G.~M. \textit{et al.} Scattering and interference in epitaxial graphene. \emph{Science} \textbf{317}, 219--222 (2007).

\bibitem{Mallet:2012ib}
Mallet,~P. \textit{et al.} Role of pseudospin in quasiparticle interferences in epitaxial graphene probed by high-resolution
scanning tunneling microscopy. \emph{Phys. Rev.
  B} \textbf{86}, 045444 (2012).

\bibitem{Leicht:2014jy}
Leicht,~P. \textit{et al.} \textit{In situ} fabrication of quasi-free-standing epitaxial graphene
nanoflakes on gold. \emph{ACS Nano} \textbf{8}, 3735--3742 (2014).

\bibitem{Simon:2011dv}
Simon,~L., Bena,~C., Vonau,~F., Cranney,~M. \& Aubel,~D. Fourier-transform scanning tunnelling spectroscopy: The possibility to obtain constant-energy maps and band dispersion using a local measurement. \emph{J. Phys. D: Appl.
  Phys.} \textbf{44}, 464010 (2011).

\bibitem{Wang:2008}
Wang,~B., Bocquet,~M.~L., Marchini,~S., Guenther,~S. \& Wintterlin,~J. Chemical origin of a graphene moir\'e overlayer on Ru(0001).
  \emph{Phys. Chem. Chem. Phys.} \textbf{10}, 3530--3534 (2008).

\bibitem{Wang:2010jw}
Wang,~B., Gunther,~S., Wintterlin,~J. \& Bocquet,~M.~L. Periodicity, work function and reactivity of graphene on Ru(0001) from first principles. \emph{New J. Phys.}
  \textbf{12}, 043041 (2010).

\bibitem{Stradi:2011be}
Stradi,~D. \textit{et al.} Role of dispersion forces in the structure of graphene monolayers on Ru surfaces. \emph{Phys. Rev. Lett.} \textbf{106}, 186102 (2011).

\bibitem{Stradi:2013dj}
Stradi,~D. \textit{et al.}  Lattice-matched versus lattice-mismatched models to describe epitaxial monolayer graphene on Ru(0001). \emph{Phys. Rev. B}
  \textbf{88}, 245401 (2013).

\bibitem{Tersoff:1985}
Tersoff,~J. \& Hamann,~D.~R. Theory of the scanning tunneling microscope. \emph{Phys. Rev. B} \textbf{31},
  805--813 (1985).

\bibitem{Blochl:1994}
Bl{\"o}chl,~P.~E. Projector augmented-wave method. \emph{Phys. Rev. B} \textbf{50}, 17953--17979 (1994).

\bibitem{Perdew:1996}
Perdew,~J., Burke,~K.\& Ernzerhof,~M. Generalized gradient approximation made simple. \emph{Phys. Rev. Lett.} \textbf{77}, 3865--3868 (1996).

\bibitem{Kresse:1994}
Kresse,~G. \& Hafner,~J. Norm-conserving and ultrasoft pseudopotentials for first-row and transition elements. \emph{J Phys.: Condens. Matter} \textbf{6},
  8245--8257 (1994).
  
\bibitem{Grimme:2006}
Grimme,~S. Semiempirical GGA-type density functional constructed with a long-range dispersion correction. \emph{J. Comput. Chem.} \textbf{27}, 1787--1799 (2006).

\bibitem{Popescu:2012bq}
Popescu,~V. \& Zunger,~A. Extracting $E$ versus $k$ effective band structure from supercell calculations on alloys and impurities. \emph{Phys. Rev. B} \textbf{85},
  085201 (2012).

\bibitem{Medeiros:2014ka}
Medeiros,~P. V.~C., Stafstr{\"o}m,~S. \& Bj{\"o}rk,~J. Effects of extrinsic and intrinsic perturbations on the electronic structure of graphene: Retaining an effective primitive cell band structure by band unfolding. \emph{Phys. Rev. B}
  \textbf{89}, 041407 (2014).
  
\end{thebibliography}
\end{document}